# A Family of Adaptive Filter Algorithms in Noise Cancellation for Speech Enhancement

Sayed. A. Hadei, *Student Member IEEE* and M. lotfizad

***Abstract***— In many application of noise cancellation, the changes in signal characteristics could be quite fast. This requires the utilization of adaptive algorithms, which converge rapidly. Least Mean Squares (LMS) and Normalized Least Mean Squares (NLMS) adaptive filters have been used in a wide range of signal processing application because of its simplicity in computation and implementation. The Recursive Least Squares (RLS) algorithm has established itself as the "ultimate" adaptive filtering algorithm in the sense that it is the adaptive filter exhibiting the best convergence behavior. Unfortunately, practical implementations of the algorithm are often associated with high computational complexity and/or poor numerical properties. Recently adaptive filtering was presented, have a nice tradeoff between complexity and the convergence speed. This paper describes a new approach for noise cancellation in speech enhancement using the two new adaptive filtering algorithms named fast affine projection algorithm and fast Euclidean direction search algorithms for attenuating noise in speech signals. The simulation results demonstrate the good performance of the two new algorithms in attenuating the noise.

## I. Introduction

It is well known that two of most frequently applied algorithms for noise cancellation [1] are normalized least mean squares (NLMS) [2]-[5] and recursive least squares (RLS) [6]-[10] algorithms. Considering these two algorithms, it is obvious that NLMS algorithm has the advantage of low computational complexity. On the contrary, the high computational complexity is the weakest point of RLS algorithm but it provides a fast adaptation rate. Thus, it is clear that the choice of the adaptive algorithm to be applied is always a tradeoff between computational complexity and fast convergence. The convergence property of the FAP and FEDS algorithms is superior to that of the usual LMS, NLMS, and affine projection (AP) algorithms and comparable to that of the RLS algorithm [11]-[14]. In these algorithms, one of the filter coefficients is updated one or more at each time instant, in order to fulfill a suitable tradeoff between convergences rate and computational complexity [15]. The performance of the proposed algorithms is fully studied through the energy conservation [16], [17] analysis used in adaptive filters and the general expressions for the steady-state mean square error and transient performance analysis were derived in [15], [18].

What we propose in this paper is the use of the FAP and FEDS algorithms in noise cancellation for speech enhancement. We compare the results with classical adaptive filter algorithm such as LMS, NLMS, AP and RLS algorithms. Simulation results show the good performance of the two algorithms in attenuating the noise. In the following we find also the optimum parameter which is used in these algorithms.

We have organized our paper as follows:

In the next section, the classical adaptive algorithms such as LMS, NLMS, AP and RLS algorithms will be reviewed. In the following the FAP algorithm in [15] and FEDS in [18] will be briefly introduced. Section 4 presents the adaptive noise cancellation setup. We conclude the paper with comprehensive set of simulation results.

Throughout the paper, the following notations are adopted:

TABLE I
NOTATIONS

| Symbol | descriptions |
|---|---|
| $\lvert . \rvert$ | Norm of a scalar |
| $\lVert . \rVert^2$ | Squared Euclidean norm of a vector |
| $(.)^T$ | Transpose of a vector or a matrix |
| $(.)^{-1}$ | Inverse of a scalar or a matrix |
| $< .,. >$ | Inner product of two vectors |

S. A. Hadei is with the School of Electrical Engineering, Tarbiat Modares University, Tehran, Iran (e-mail: a.hadei@modares.ac.ir).

M. Lotfizad is with the School of Electrical Engineering, Tarbiat Modares University, Tehran, P.O.Box 14115-143, Tehran, Iran. (e-mail: lotfizad@modares.ac.ir).

## II. Background on LMS, NLMS, APA and RLS Algorithms

In Fig. 1, we show the prototypical adaptive filter setup, where $x(n)$, $d(n)$ and $e(n)$ are the input, the desired and the output error signals, respectively. The vector $\underline{h}(n)$ is the $M \times 1$ column vector of filter coefficient at time $n$, in such a way that the output of signal, $y(n)$, is good estimate of the desired signal, $d(n)$.

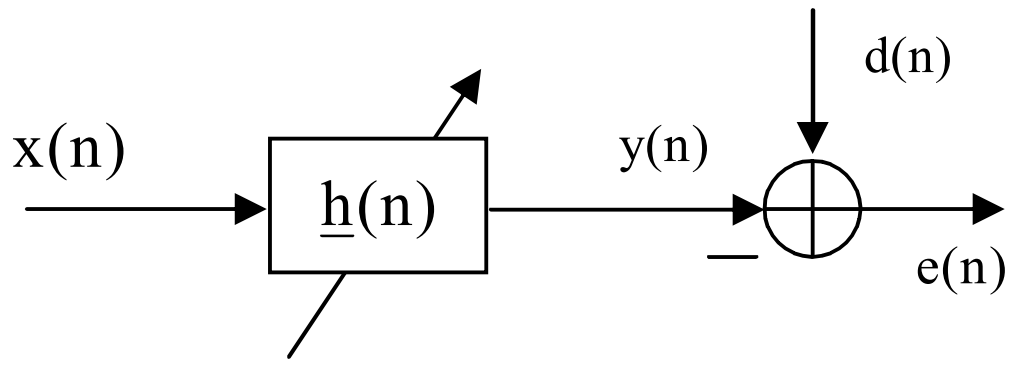

Fig.1. Prototypical adaptive filter setup

It is well known that the filter vector update equation for the LMS algorithm is given by [9]:

$$\underline{h}(n+1) = \underline{h}(n) + \mu\, \underline{x}(n) e(n), \tag{1}$$

where

$$\underline{x}(n) = [x(n), x(n-1), \ldots, x(n-M+1)]^T , \tag{2}$$

and $\mu$ is the step-size that determines the convergence speed and steady-state mean-square error (MSE). Also, the output error signal, $e(n)$, is given by

$$e(n) = d(n) - \underline{h}^T(n) \underline{x}(n). \tag{3}$$

To increase the convergence speed of the LMS algorithm, the NLMS and AP algorithms was proposed which can be stated as [9]

$$\underline{h}(n+1) = \underline{h}(n) + \frac{\mu}{\left\| \underline{x}(n) \right\|^2} \underline{x}(n) e(n) \tag{4}$$

$$\underline{h}(n+1) = \underline{h}(n) + \mu X^T(n) (\varepsilon I + X(n) X^T(n))^{-1} [\underline{d}(n) - X(n) \underline{h}(n)] \tag{5}$$

where

$$X(n) = [\underline{x}(n), \underline{x}(n-1), \ldots, \underline{x}(n-K+1)]^T \tag{6}$$

and

$$\underline{d}(n) = [d(n), d(n-1), \ldots, d(n-K+1)]^T \tag{7}$$

The filter vector update equation in RLS algorithm is [14]:

$$\underline{h}(n+1) = \underline{h}(n) + C^{-1}(n) \underline{x}(n) e(n), \tag{8}$$

where $C(n)$ is the estimation of the autocorrelation matrix. This matrix is given by

$$C(n) = \sum_{i=0}^{n} \lambda^{n-i} \underline{x}(i) \underline{x}^T(i) . \tag{9}$$

The $\lambda$ parameter is the forgetting factor and $0 << \lambda < 1$.

## III. FAPA and FEDS Algorithms

### *A. Notation and problem description*

With reference to Figure 1, the error signal, $e(n)$, can be expressed as:

$$e(n) = d(n) - \sum_{k=\circ}^{M-1} h_k(n) x(n-k). \tag{10}$$

Considering the samples $n-L+1, n-L+2, \ldots, n,$ where we focus on the situation where $L > M$, Eq.7 can be written as:

$$\underline{e}(n) = \underline{d}(n) - X(n) \underline{h}(n), \tag{11}$$

where

$$X(n) = [\underline{x}_\circ(n), \underline{x}_1(n), \ldots, \underline{x}_{M-1}(n)]. \tag{12}$$

These columns are furthermore defined through

$$\underline{x}_j(n) = [x(n-j), x(n-j-1), \ldots, x(n-j-L+1)]^T . \tag{13}$$

The vector of desired signal samples is given by

$$\underline{d}(n) = [d(n), d(n-1), \ldots, d(n-L+1)]^T , \tag{14}$$

and $\underline{e}(n)$ is defined similarly. The adaptive filtering problem can now be formulated as the task of finding the update for $\underline{h}(n)$, at each time instant $n$, such that the error is made as small as possible.

Note that $X(n)\underline{h}(n)$ can be written as

$$X(n)\underline{h}(n) = \sum_{k=0}^{M-1} h_k(n)\underline{x}_k(n), \tag{15}$$

i.e. as a weighted sum of the columns of $X(n)$ with the elements of $\underline{h}(n)$ being the weighting factors. A greedy algorithm for successively building (better) approximations to a given vector using linear combinations of vectors from a given set is the BMP algorithm. Inspired by this algorithm, conceived and developed in another context and with other motivations than those of this paper, we devise a procedure for recursively building an approximation to $\underline{d}(n)$ using linear combinations of the columns of $X(n)$.

### B. Algorithm development

Assuming that we have an approximation to $\underline{d}(n-1)$ at time $n-1$ given by $X(n-1)\underline{h}(n-1)$, the *apriori* approximation error at time $n$ is

$$\underline{e}_0(n) = \underline{d}(n) - X(n)\underline{h}(n-1). \tag{16}$$

In building a better approximation through the update of only one coefficient in $\underline{h}(n-1)$, we would write the new error as

$$\underline{e}_1(n) = \underline{d}(n) - (X(n)\underline{h}(n-1) + X(n)h^{update}_{j_0(n)}(n)\underline{u}_{j_0(n)}) \tag{17}$$

Note that $j_0(n)$ is the index of the coefficient to be update in the zero'th P-iteration at time $n$, and $\underline{u}_j$ is the M-vector with 1 in position $j$ and 0 in all other positions. Intuitively, it would make sense to select $j_0(n)$ as the index corresponding to that column of $X(n)$ that is most similar to the apriori approximation error of Eq. 13. Thus, coefficient $j(n)$ has been identified as the one to update. We have identified two ways of selecting $j(n)$ : I) incrementing $j(n)$ sequently by $n$ modulo $M$ and II) selecting $j(n)$ in such a way as to maximally reduce the residual of the corresponding update computation. The former selection in conjunction with Eq.14 is the FEDS algorithm, whereas the latter in conjunction with Eq.14 results in the FAP algorithm. Thus, in the FAPA, $j_0(n)$ is found as the index of the column of $X(n)$ onto which $\underline{e}_0(n)$ has its maximum projection, -or in other words:

$$j_0(n) = \arg\max_j \frac{\left|\langle \underline{e}_0(n), \underline{x}_j(n) \rangle\right|}{\left\|\underline{x}_j(n)\right\|}, \tag{18}$$

Where $<.,.>$ denotes an inner product between the two vector arguments. Given the index $j_0(n)$, the update of the corresponding filter coefficient is

$$h_{j_0(n)}(n) = h_{j_0(n)}(n-1) + h^{update}_{j_0(n)}(n), \tag{19}$$

where $h^{update}_{j_0(n)}(n)$ is the value of the projection of $\underline{e}_0(n)$ onto the unit vector with direction given by $\underline{x}_{j_0(n)}(n)$, i.e.:

$$h^{update}_{j_0(n)}(n) = \frac{<\underline{e}_0(n), \underline{x}_{j_0(n)}(n)>}{\left\|\underline{x}_{j_0(n)}(n)\right\|^2}. \tag{20}$$

Thus, the zero'th P-iteration updates the filter vector as follows:

$$\underline{h}^{(0)}(n) = \underline{h}(n-1) + h^{update}_{j_0(n)}(n)\underline{u}_{j_0(n)}. \tag{21}$$

To have control on the convergence speed and stability of the algorithms, we introduce the step-size in the algorithm as following:

$$\underline{h}^{(0)}(n) = \underline{h}(n-1) + \mu h^{update}_{j_0(n)}(n)\underline{u}_{j_0(n)} \tag{22}$$

Given this, the updated error expression of Eq.14 can be written as:

$$\underline{e}_1(n) = \underline{d}(n) - X(n)\underline{h}^{(0)}(n). \tag{23}$$

If we want to do more than one P-iteration at time $n$, the procedure described above starting with finding the maximum projection of $\underline{e}_0(n)$ onto a column of $X(n)$ can be repeated with $\underline{e}_1(n)$ taking the role of $\underline{e}_0(n)$. This can be repeated as many times as desired, say P times, leading to a sequence of coefficient updates:

$$h_{j_{\circ}(n)}(n), h_{j_1(n)}(n), \ldots, h_{j_{P-1}(n)}(n). \qquad (24)$$

Note that if $P > 2$ it is entirely possible that one particular coefficient is updated more than once at a given time $n$. The resulting filter coefficient vector after P iterations at time $n$ is denoted $\underline{h}^{(P-1)}(n)$, but where there is no risk of ambiguity, we shall refer to this filter vector simply as $\underline{h}(n)$.

The procedure described above corresponds to applying the BMP algorithm to a dictionary of vectors given by the columns of $X(n)$ for the purpose of building an approximation to $\underline{d}(n)$. The only difference is that we do this for each new time instant $n$ while keeping the results of the BMP from the previous time instant $n-1$. It is interesting to note that a slightly different, but equivalent, procedure to the one described above would result if we tried to find the least squares solution to the over determined set of equations (remember $L > M$):

$$X(n)\underline{h}(n) = \underline{d}(n) \qquad (25)$$

Subject to the constrain that, given an initial solution, say $\underline{h}_{\circ}(n)$, we are allowed to adjust only one element of this vector.

From the above, it is evident that the key computations of our adaptive filter algorithm are those of Eqs.15 and 17. Making use of Eqs. 13 and 12, we find

$$j_{\circ}(n) = \arg\max_{j} \frac{1}{\left\| \underline{x}_j(n) \right\|} \Big| < \underline{d}(n), \underline{x}_j(n) > - \sum_{k=\circ}^{M-1} h_k(n-1) < \underline{x}_k(n), \underline{x}_j(n) > \Big| , \qquad (26)$$

and

$$h_{j_\circ(n)}^{update}(n) = \frac{1}{\left\| \underline{x}_{j_\circ(n)}(n) \right\|^2} \{ < \underline{d}(n), \underline{x}_{j_\circ(n)}(n) > - \sum_{k=\circ}^{M-1} h_k(n-1) < \underline{x}_k(n), \underline{x}_{j_\circ(n)}(n) > \} . \qquad (27)$$

These are the pertinent equations if one coefficient update, i.e. one P-iteration is performed for each new signal sample. Note that having computed the terms of Eq. 23, very little additional work is involved in finding the update of Eq. 24. It is instructive to explicitly state these equations also for iteration no. $i > 0$ at time $n$:

$$j_i(n) = \arg\max_{j} \frac{1}{\left\| \underline{x}_j(n) \right\|} \Big| < \underline{d}(n), \underline{x}_j(n) > - \sum_{k=\circ}^{M-1} h_k^{(i-1)}(n) < \underline{x}_k(n), \underline{x}_j(n) > \Big| \qquad (28)$$

and

$$h_{j_i(n)}^{update}(n) = \frac{1}{\left\| \underline{x}_{j_i(n)}(n) \right\|^2} \{ < \underline{d}(n), \underline{x}_{j_i(n)}(n) > - \sum_{k=\circ}^{M-1} h_k^{(i-1)}(n) < \underline{x}_k(n), \underline{x}_{j_i(n)}(n) > \} . \qquad (29)$$

From these equations it is evident that some terms depend only on $n$, i.e. they need to be computed once for each $n$ and can subsequently be used unchanged for all P-iterations at time $n$. Other terms depend on both $n$ and the P-iteration index and must consequently be updated for each P-iteration. Since we must associate the update depending only on $n$ with iteration no. 0, this is the computationally most expensive update.

From the above it is evident that the inner products $< \underline{d}(n), \underline{x}_j(n) >$ and $< \underline{x}_k(n), \underline{x}_j(n) >$ play prominent roles in the computations involved in the algorithm. As formulated up to this point, obvious recursions for these inner products are

$$< \underline{d}(n), \underline{x}_j(n) > = < \underline{d}(n-1), \underline{x}_j(n-1) > + d(n)x(n-j) - d(n-L)x(n-j-L), \qquad (30)$$

and

$$< \underline{x}_k(n), \underline{x}_j(n) > = < \underline{x}_k(n-1), \underline{x}_j(n-1) > + x(n-k)x(n-j) - x(n-k-L)x(n-j-L) \qquad (31)$$

We close this section by pointing out that efficient implementations of FEDS/FAP are available. For exponentially weighted and sliding window versions, it is known that implementations having a multiplicative complexity given by $(5+P)M$ can be devised [15]. If we use a block exponentially weighted version [19], implementations with a multiplicative complexity of $(3+P)M$ are possible.

## IV. Adaptive noise Cancellation

Fig. 2 shows the adaptive noise cancellation setup. In this application, the corrupted signal passes through a filter that tends to suppress the noise while leaving the signal unchanged. This process is an adaptive process, which means it cannot require a priori knowledge of signal or noise characteristics.

Adaptive noise cancellation algorithms utilize two or more microphones (sensor). One microphone is used to measure the speech + noise signal while the other is used to measure the noise signal alone. The technique adaptively adjusts a set of filter coefficients so as to remove the noise from the noisy signal. This technique, however, requires that the noise component in the corrupted signal and the noise in the reference channel have high coherence. Unfortunately this is a limiting factor, as the microphones need to be separated in order to prevent the speech being included in the noise reference and thus being removed. With large separations the coherence of the noise is limited and this limits the effectiveness of this technique. In summary, to realize the adaptive noise cancellation, we use two inputs and an adaptive filter. One input is the signal corrupted by noise (Primary Input, which can be expressed as $s(n)+n_0(n)$). The other input contains noise related in some way to that in the main input but does not contain anything related to the signal (Noise Reference Input, expressed as $n_1(n)$). The noise reference input pass through the adaptive filter and output $y(n)$ is produced as close a replica as possible of $n_0(n)$. The filter readjusts itself continuously to minimize the error between $n_0(n)$ and $y(n)$ during this process. Then the output $y(n)$ is subtracted from the primary input to produce the system output $e = s + n_0 - y$, which is the denoised signal. Assume that $s, n_0$, $n_1$ and $y$ are statistically stationary and have zero means. Suppose that $s$ is uncorrelated with $n_0$ and $n_1$, but $n_1$ is correlated with $n_0$. We can get the following equation of expectations:

$$E[e^2] = E[s^2] + E[(n_0 - y)^2] \tag{32}$$

When the filter is adjusted so that $E[e^2]$ is minimized, $E[(n_0 - y)^2]$ is also minimized. So the system output can serve as the error signal for the adaptive filter. The adaptive noise cancellation configuration is shown in Fig. 2. In this setup, we model the signal path from the noise source to primary sensor as an unknown FIR channel $W_e$. Applying the adaptive filter to reference noise at reference sensor, we then employ an adaptive algorithm to train the adaptive filter to match or estimate the characteristics of unknown channel $W_e$.

If the estimated characteristics of unknown channel have negligible differences compared to the actual characteristics, we should be able to successfully cancel out the noise component in corrupted signal to obtain the desired signal. Notice that both of the noise signals for this configuration need to be uncorrelated to the signal $s(n)$. In addition, the noise sources must be correlated to each other in some way, preferably equal, to get the best results.

Do to the nature of the error signal, the error signal will never become zero. The error signal should converge to the signal $s(n)$, but not converge to the exact signal. In other words, the difference between the signal $s(n)$ and the error signal $e(n)$ will always be greater than zero. The only option is to minimize the difference between those two signals.

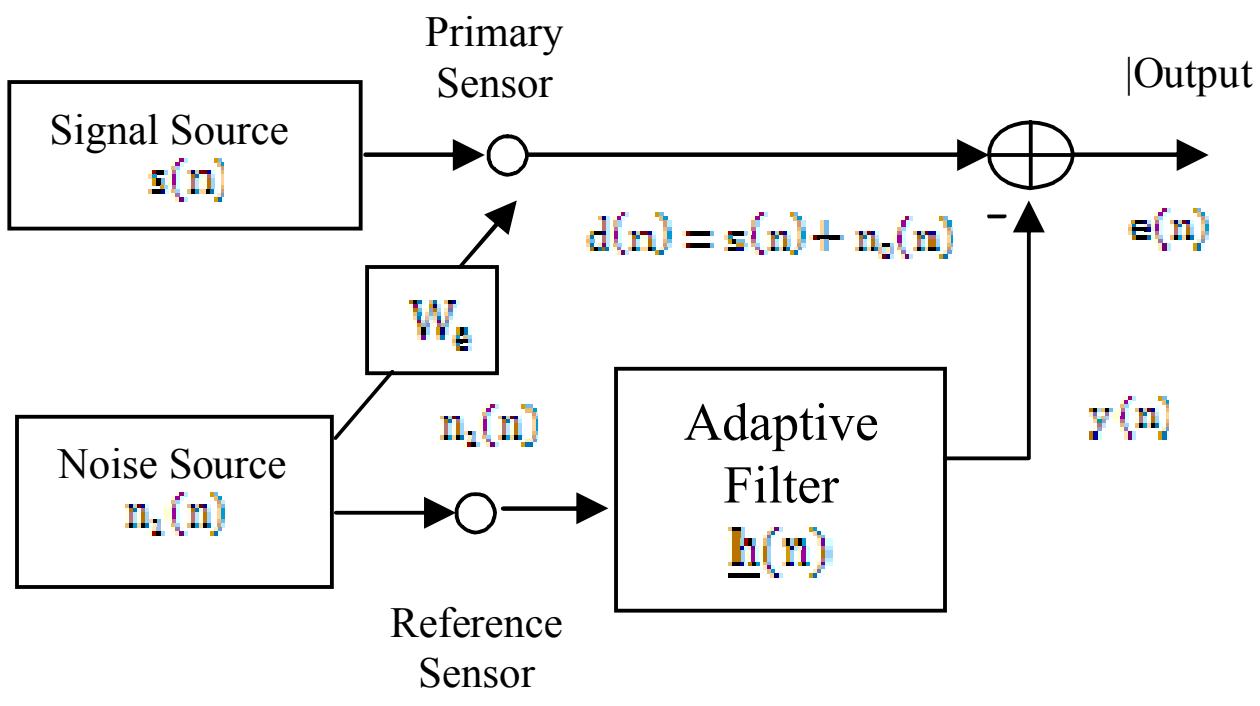

Fig. 2. Adaptive noise cancellation setup

## V. EXPERIMENTAL RESULTS

In this section we evaluate the performance of each algorithm in noise cancellation setup as shown in Fig. 2. The original, primary, and reference signals are from the reference [20]. The original speech is corrupted with office noise. The signal to noise ratio (SNR) of the primary signal is -10.2180 dB. This signal is then processed as in Fig. 2. Fig. 3 shows the signals.

The order of the filter was set to M=8. The parameter $\mu$ was set to 0.002 in the LMS and 0.005 in the NLMS and AP algorithms. Fig. 4 shows the filtered output signal and the mean squared error (learning curve) in the LMS algorithm. The SNR of the filtered signal is calculated for this experiment. The SNR improvement (SNRI) is defined as the final SNR minus the original SNR. The SNRI in the LMS algorithm is 13.5905. Fig. 5, 6 shows the results for NLMS and AP algorithms. As we can see the convergence speed in the NLMS and AP algorithms is faster than LMS algorithm. This fact can be seen in both filtered output and learning curve. For the NLMS and AP algorithms the SNRI are respectively 16.8679, 20.0307.

In Figs. 7-8, we presented the results for FEDS and FAP algorithms. The parameters was set to $L = 25, P = 8, \mu = 0.002$.

The results show that the FEDS and FAP has faster convergence speed than LMS, NLMS, AP algorithms and comparable with the RLS algorithm. The SNRI in these algorithms is 22.2623 and 24.9078.
Fig. 9 shows the results for RLS algorithm. In this algorithm, the parameter λ was set to 0.99. The results show that the RLS algorithm has faster convergence speed compared with LMS, NLMS and AP algorithms. The SNRI in this algorithm is 29.7355. Table 2 summarizes the SNRI results.

Figs. 10-15 show the filter coefficients evolutions of the, LMS, NLMS, AP, FEDS, FAP and RLS algorithms. Again, the results show that the performance of the FEDS and FAP is better than the LMS, NLMS and AP algorithms and comparable with the RLS algorithm.

In order to obtain the optimum order the filter in FEDS and FAP algorithms, we changed the order of filter from 1 to 300 and then calculated SNRI for each order of filter. Figs. 16-17. is SNRI versus order of filter. In this simulation the parameters was set to $L = 25, P = 1, \mu = 0.002$ This figure shows that the FEDS and FAP has the maximum SNRI in $M = 8$. Figs. 18-19 shows the SNRI versus $L$. The parameters was set to $M = 8, P = 1, \mu = 0.002$ This figure shoes that the FEDS and FAP has the maximum SNRI for $L = 25$. Figs. 20-21 and 22-23 show the SNRI versus the $\mu$, and $P$ respectively.

TABLE II
SNR IMPROVEMENT IN DB

| Algorithm | SNRI(db) |
|---|---|
| *LMS* | 13.5905 |
| *NLMS* | 16.8679 |
| *APA* | 20.0307 |
| *FEDS* | 22.2623 |
| FAPA | 24.9078 |
| RLS | 29.7355 |

## VI. CONCLUSION

In this paper we have applied FEDS and FAP algorithms on adaptive noise cancellation setup. The simulation results were compared with the classical adaptive filters, such as LMS, NLMS, AP and RLS algorithms, for attenuating noise in speech signals. In each algorithm the time evolution of filter taps, mean square error, and the output of filter were presented. The simulation results show that the convergence rate of these algorithms is comparable with the RLS algorithm. Also, the optimum values of the FEDS and FAP algorithms were calculated through experiments. In these algorithms, the number of iterations to be performed at each new sample time is a user selected parameter giving rise to attractive and explicit tradeoffs between convergence/tracking properties and computational complexity.

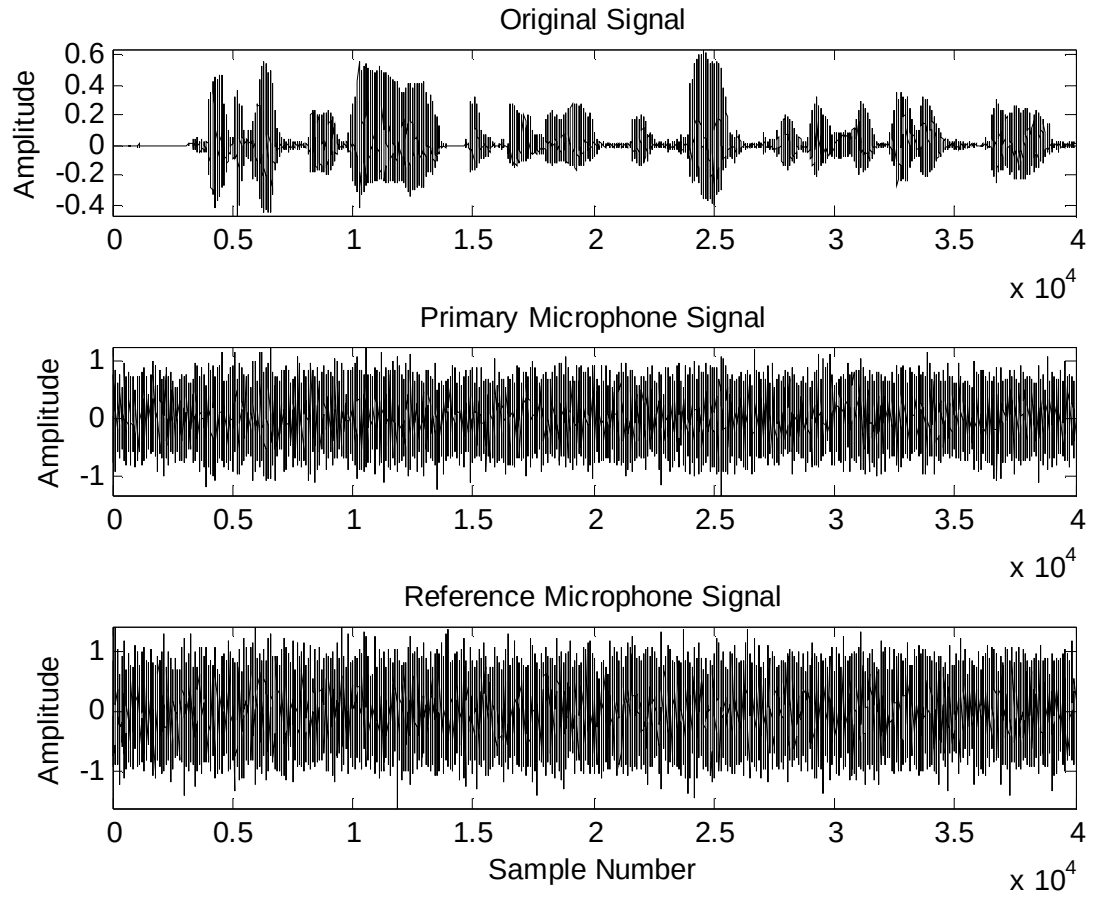

Fig. 3. Original, primary and reference signals.

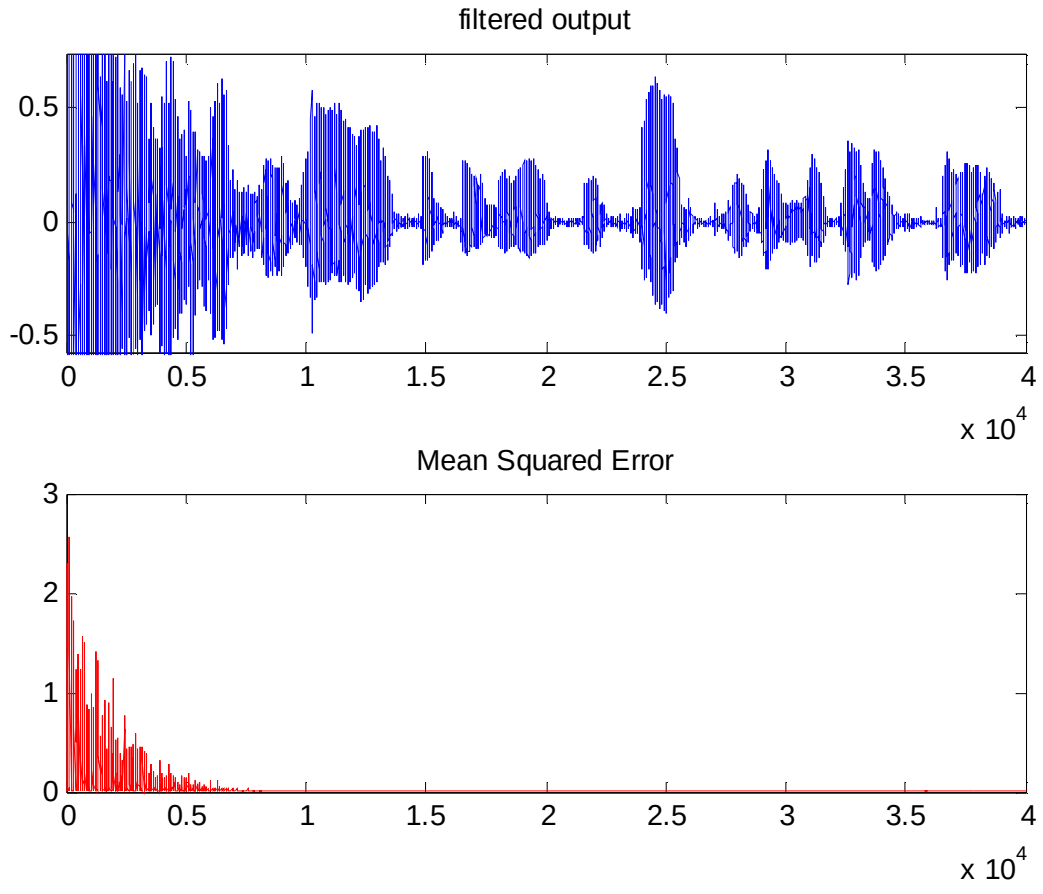

Fig. 4. Filtered output signal and MSE curve of the LMS algorithm.

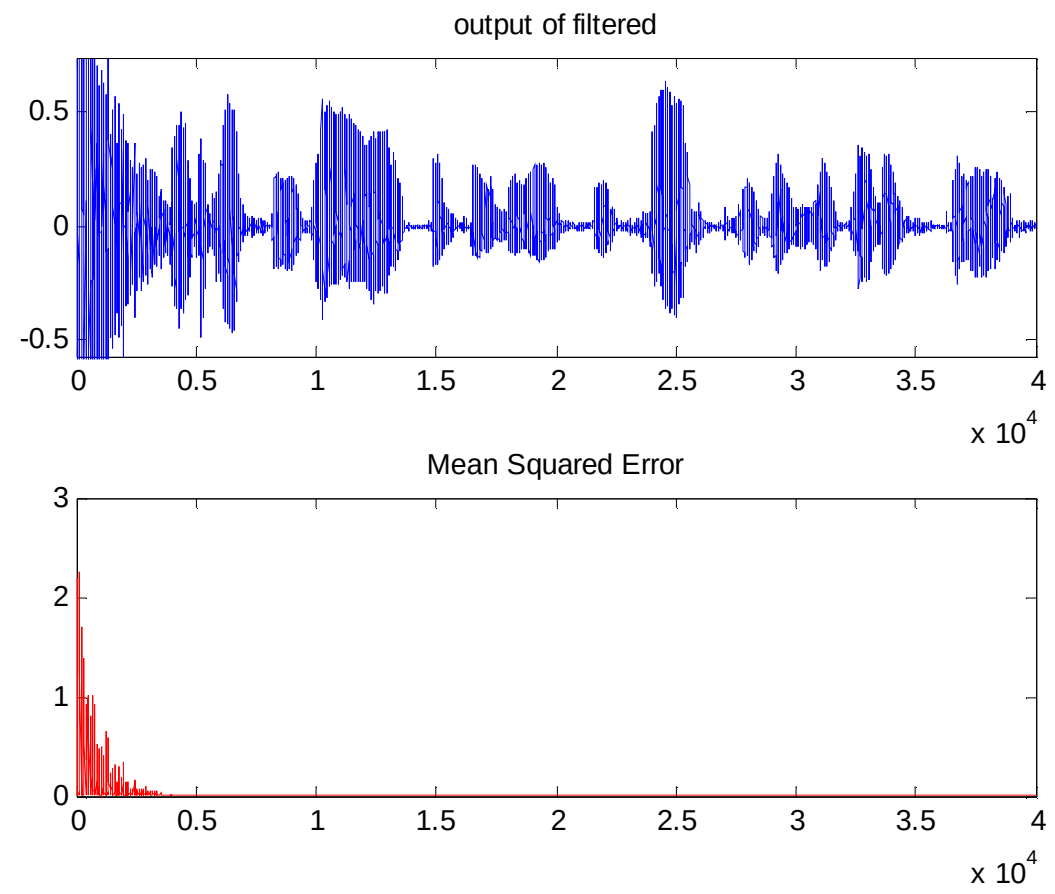

Fig. 5. Filtered output signal and MSE curve of the NLMS algorithm.

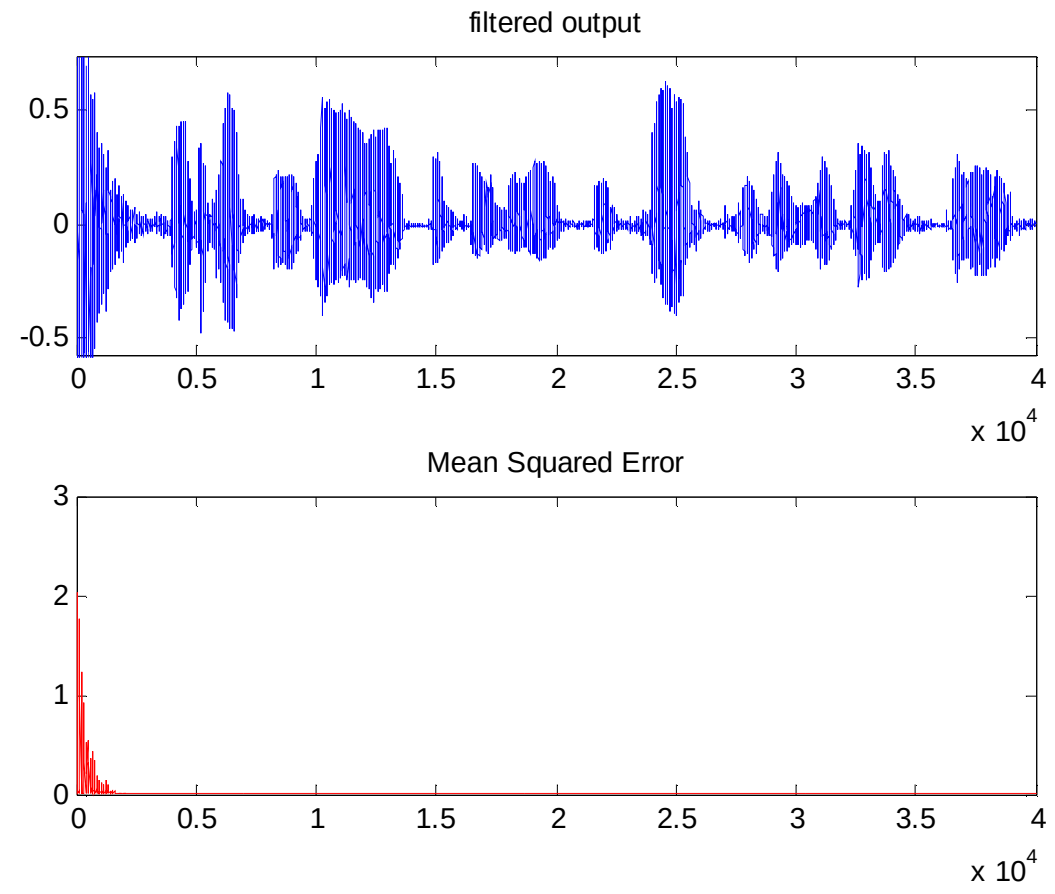

Fig. 6. Filtered output signal and MSE curve of the AP algorithm.

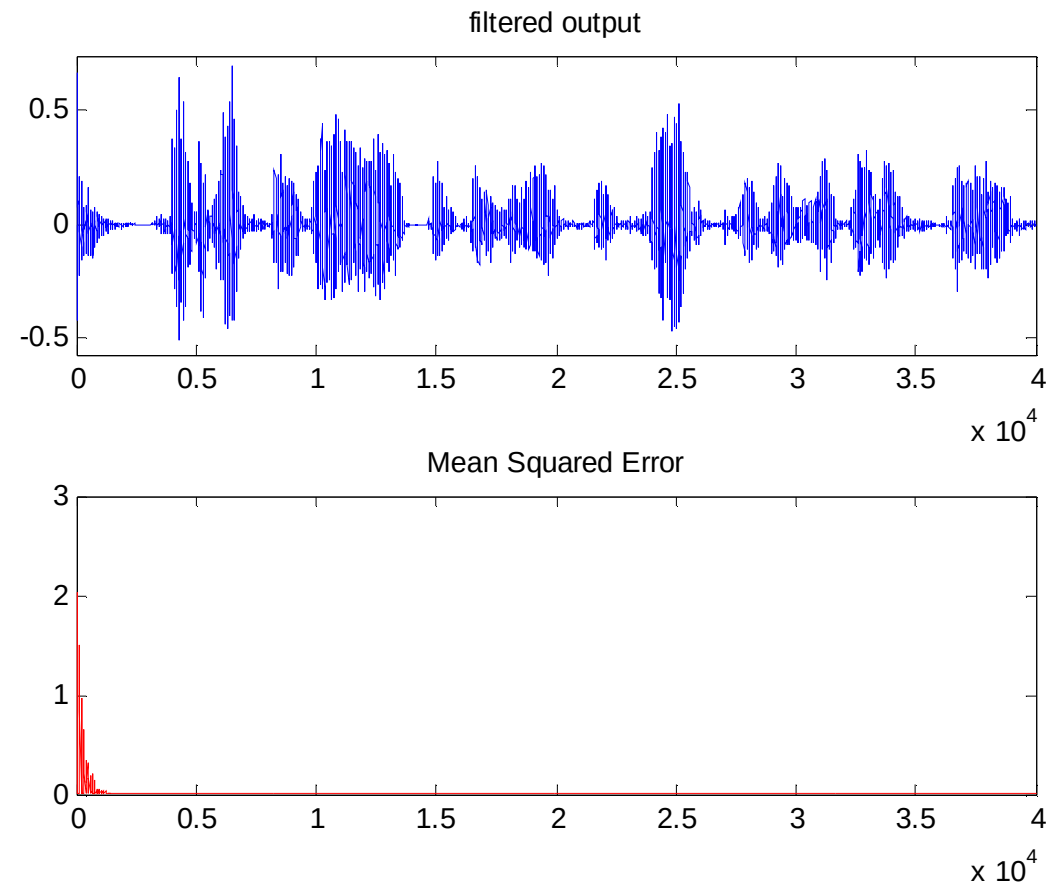

Fig. 7. Filtered output signal and MSE curve of the FEDS algorithm.

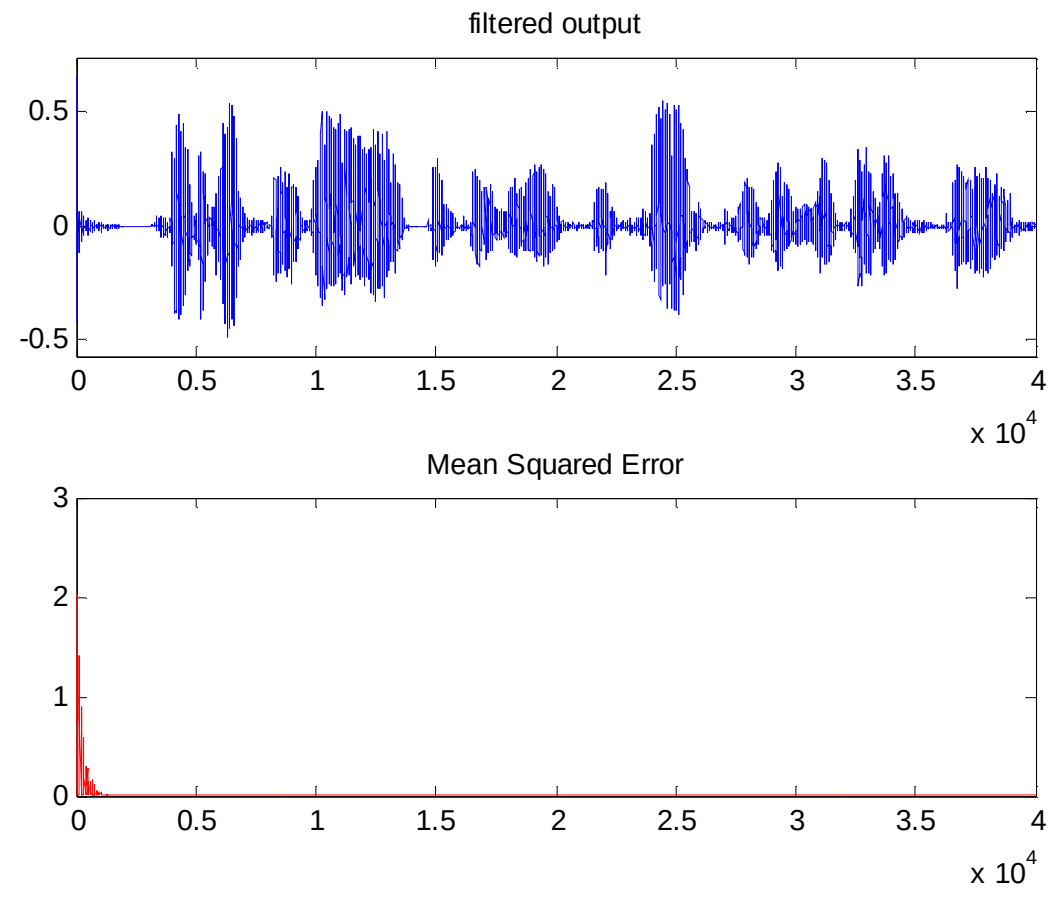

Fig. 8. Filtered output signal and MSE curve of the FAP algorithm.

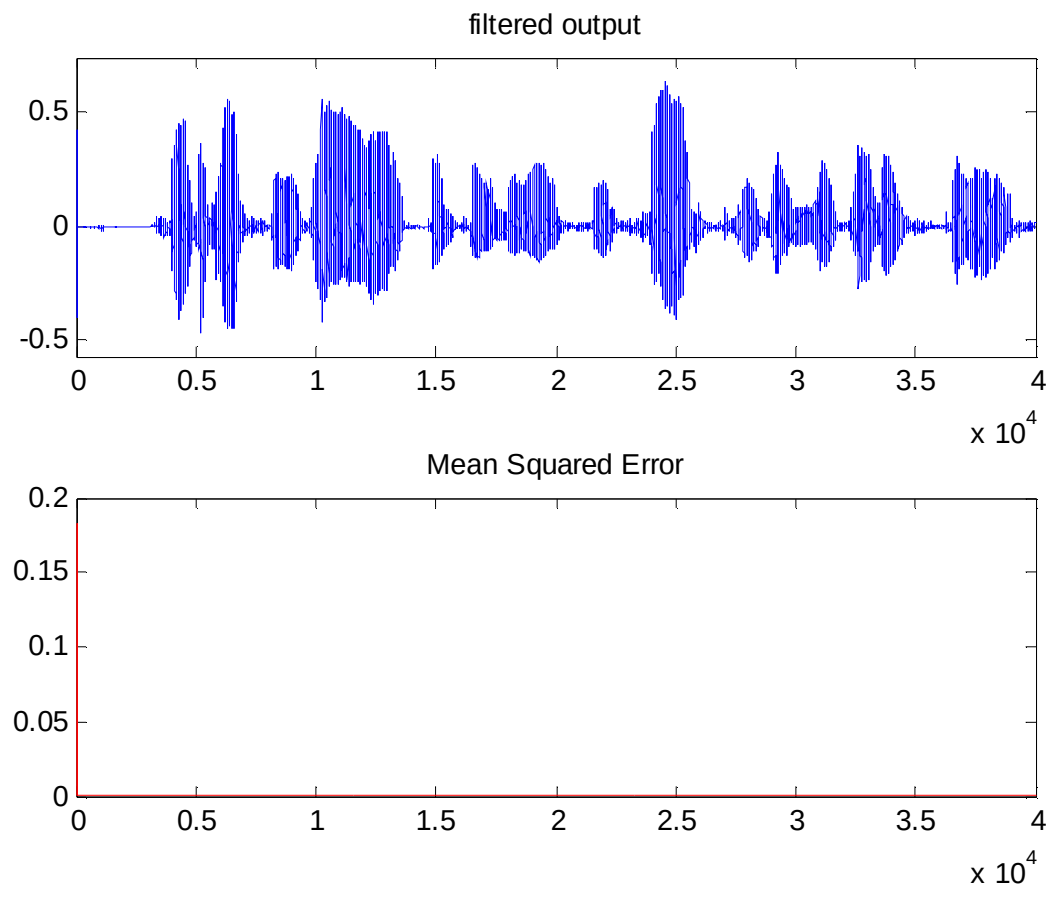

Fig. 9. Filtered output signal and MSE curve of the RLS algorithm.

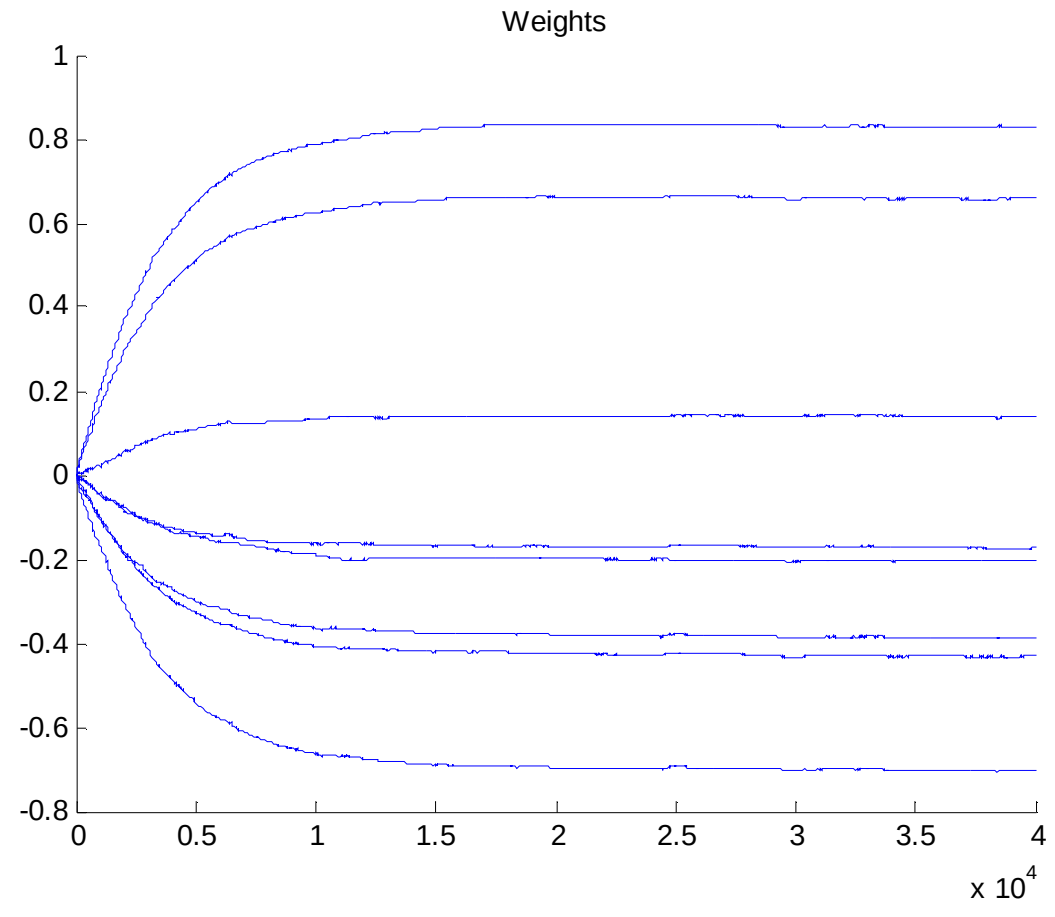

Fig. 10. Time evolution of filter taps in ANC through LMS algorithm.

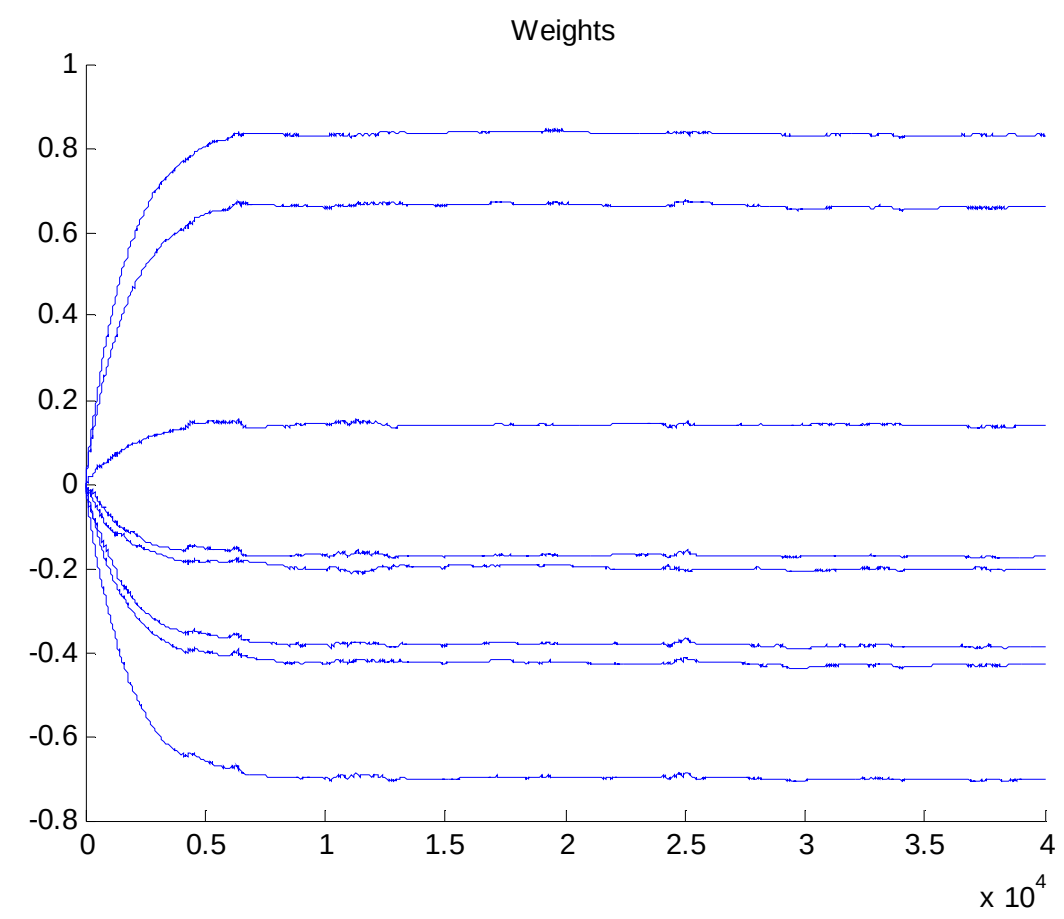

Fig. 11. Time evolution of filter taps in ANC through NLMS algorithm.

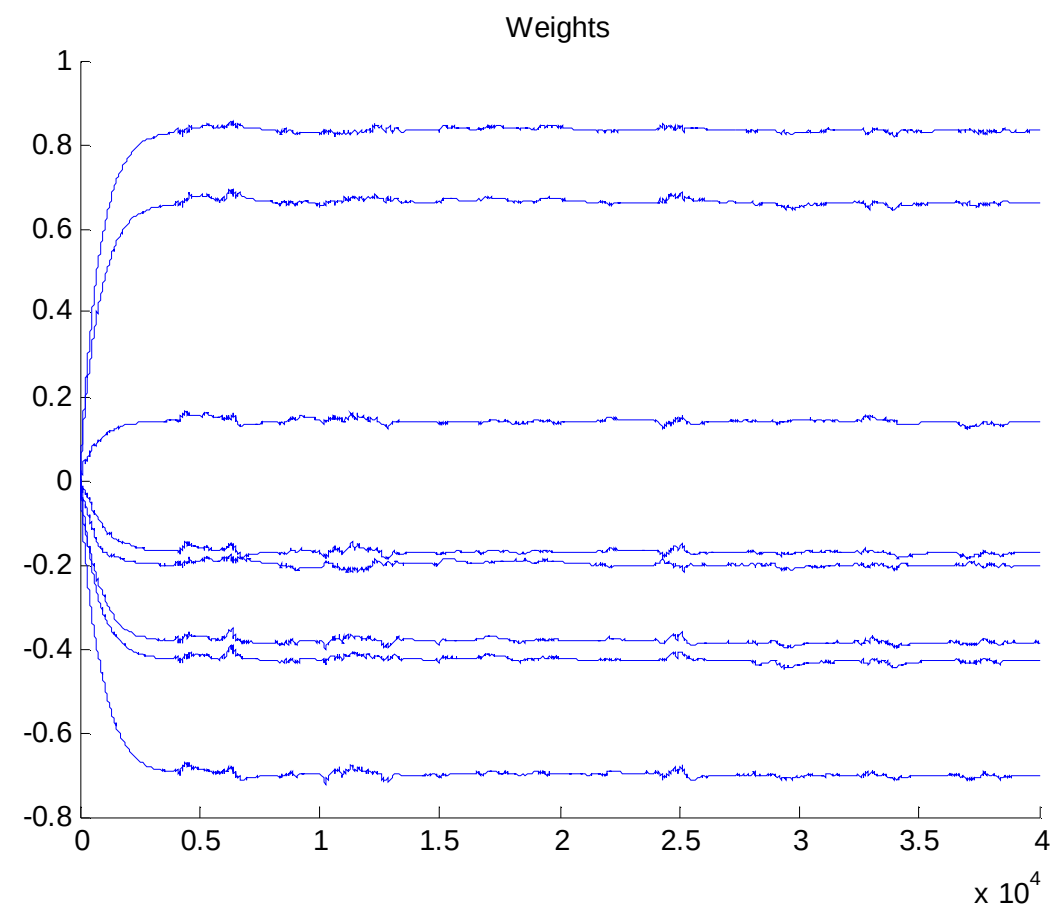

Fig. 12. Time evolution of filter taps in ANC through AP algorithm.

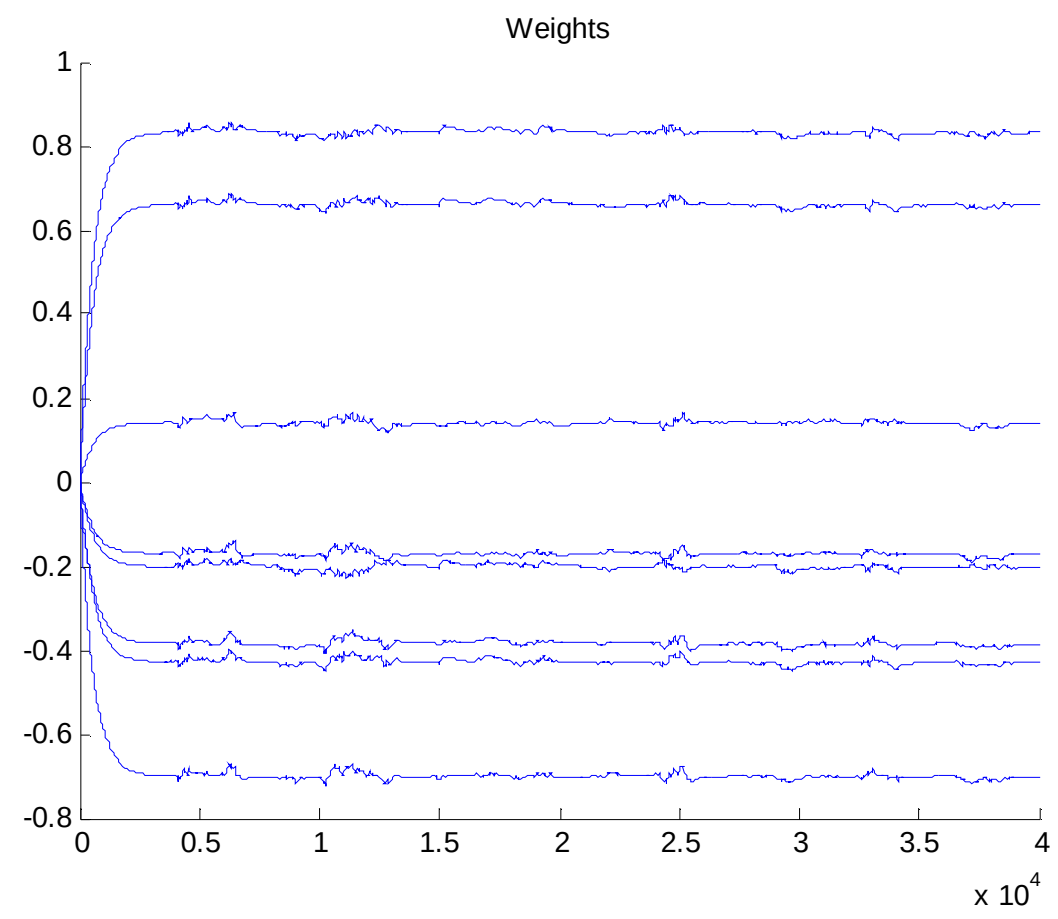

Fig. 13. Time evolution of filter taps in ANC through FEDS algorithm.

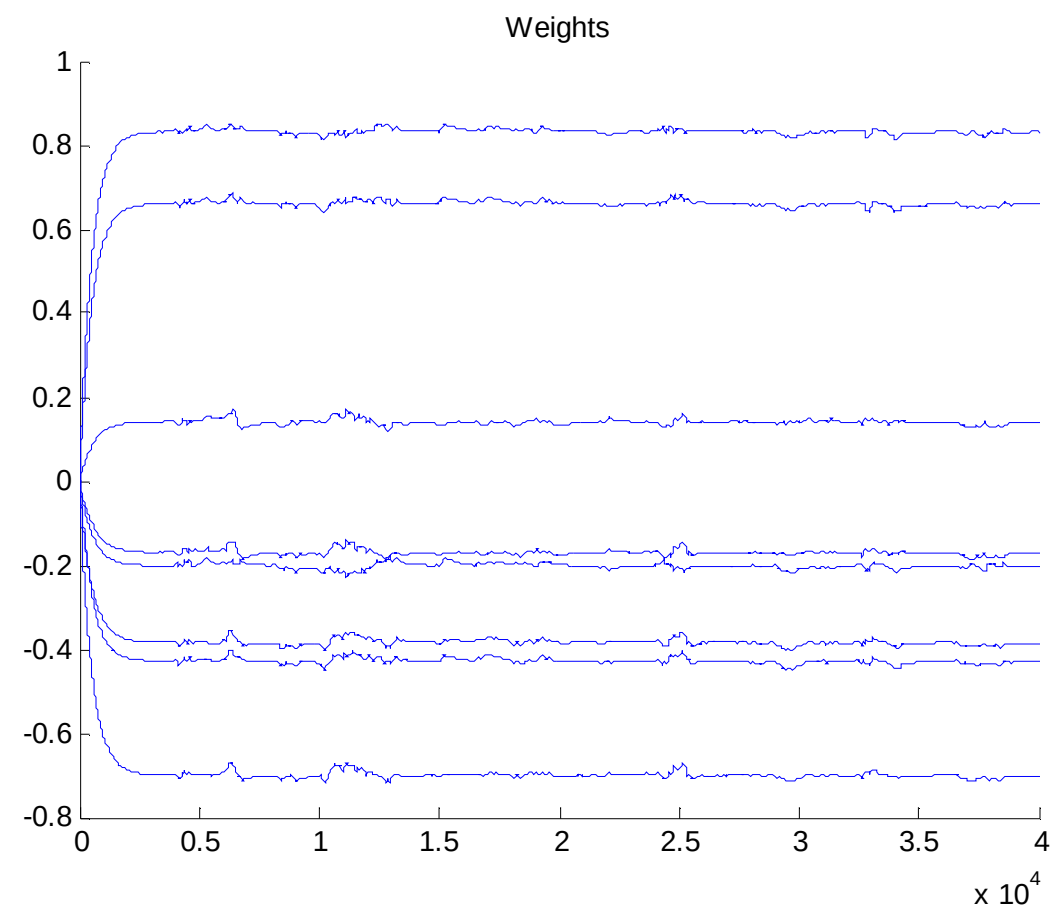

Fig. 14. Time evolution of filter taps in ANC through FAP algorithm.

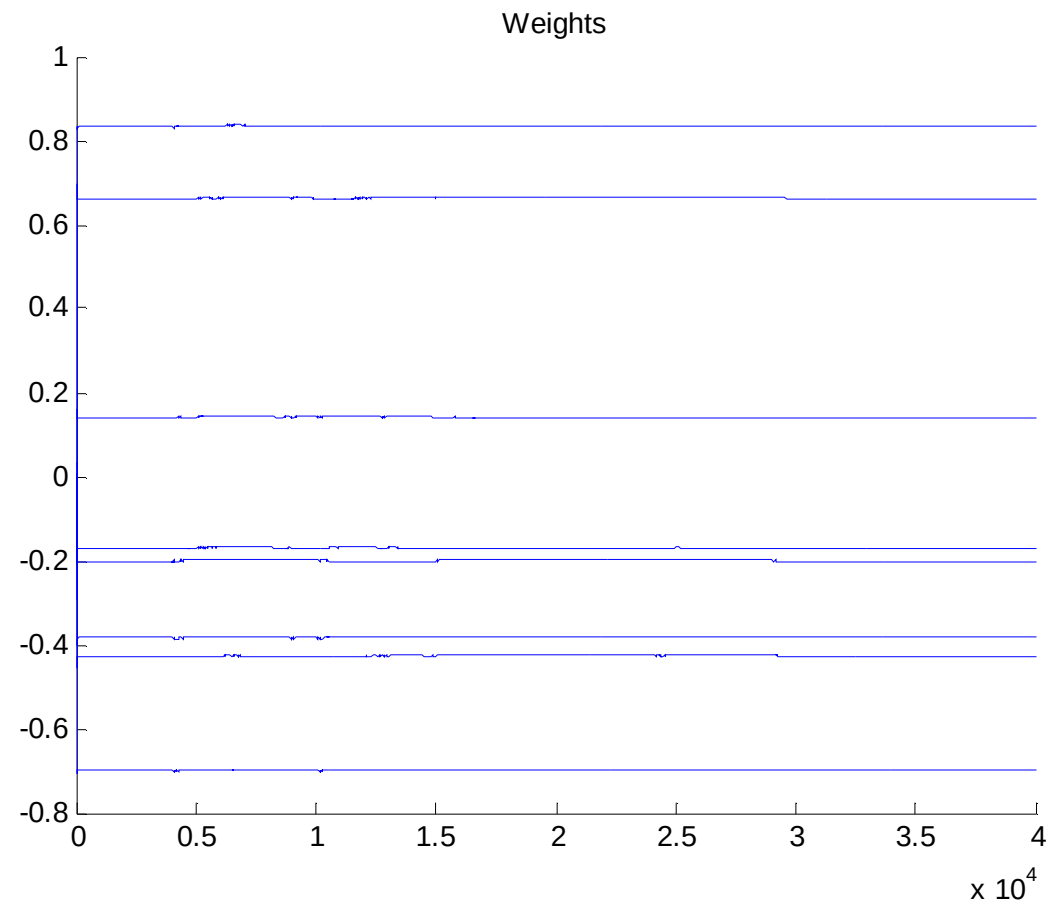

Fig. 15. Time evolution of filter taps in ANC through RLS algorithm.

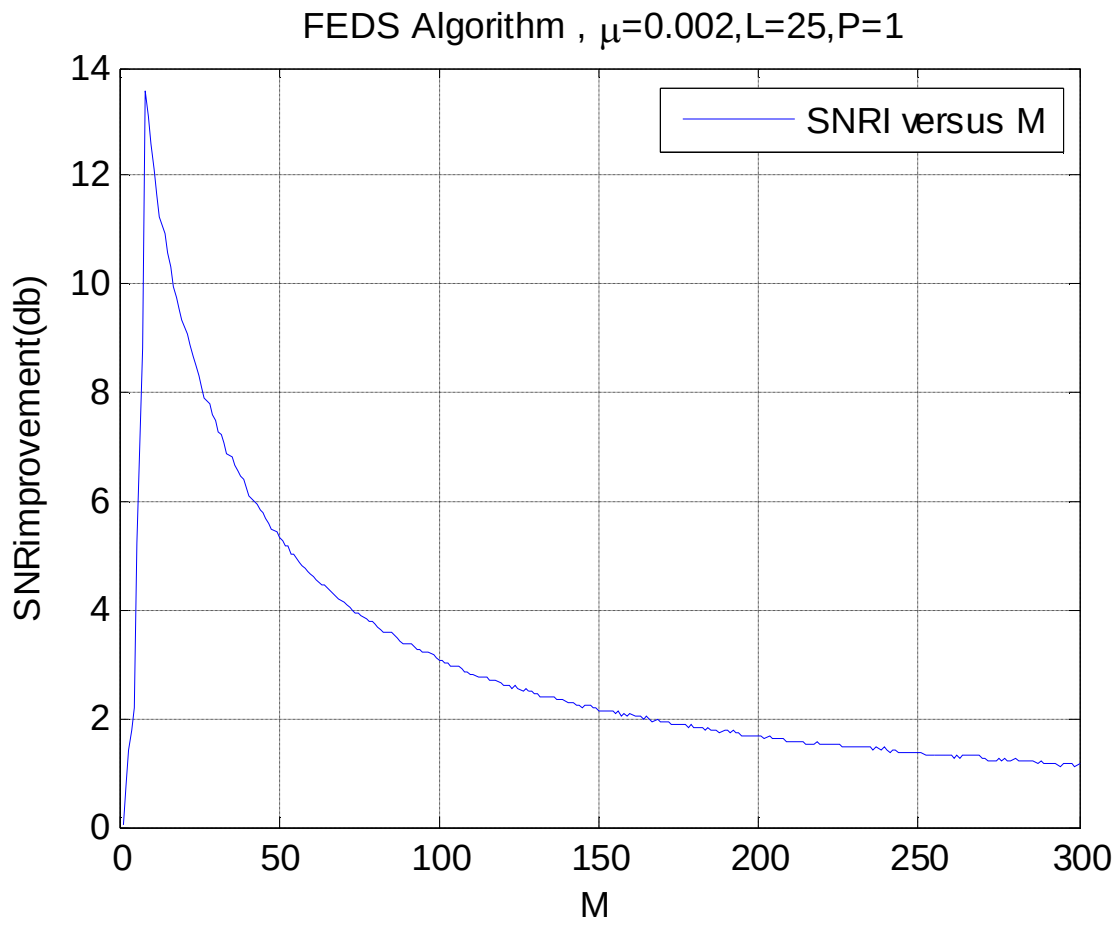

Fig. 16. SNRI versus M for FEDS algorithm.

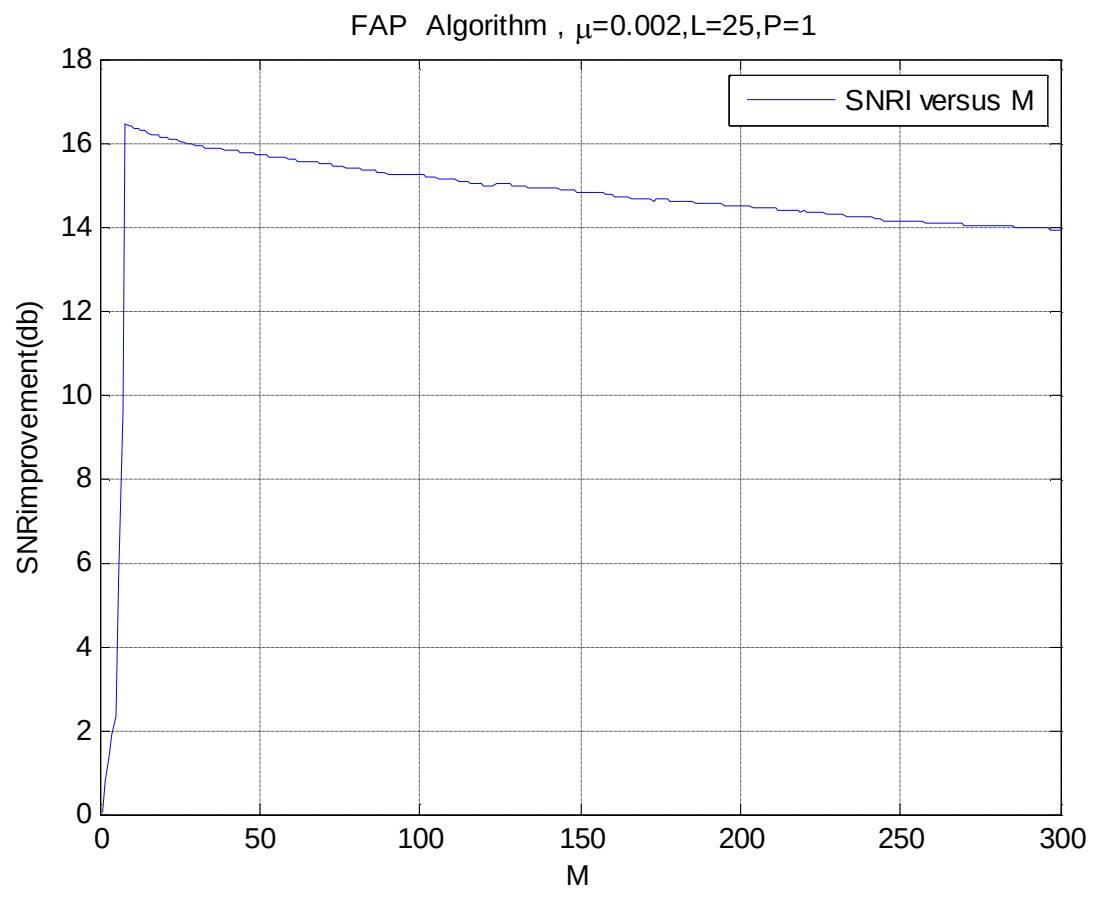

Fig. 17. SNRI versus M for FAP algorithm.

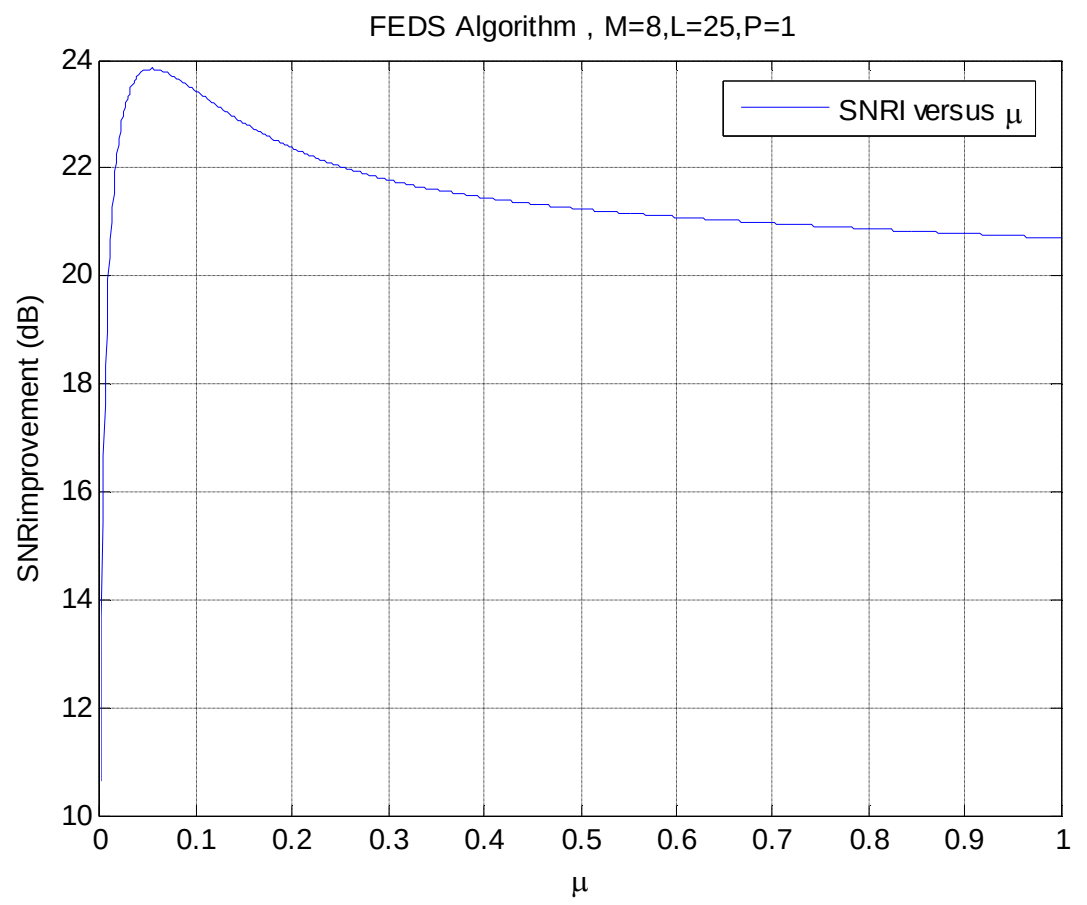

Fig. 20. SNRI versus μ for FEDS algorithm.

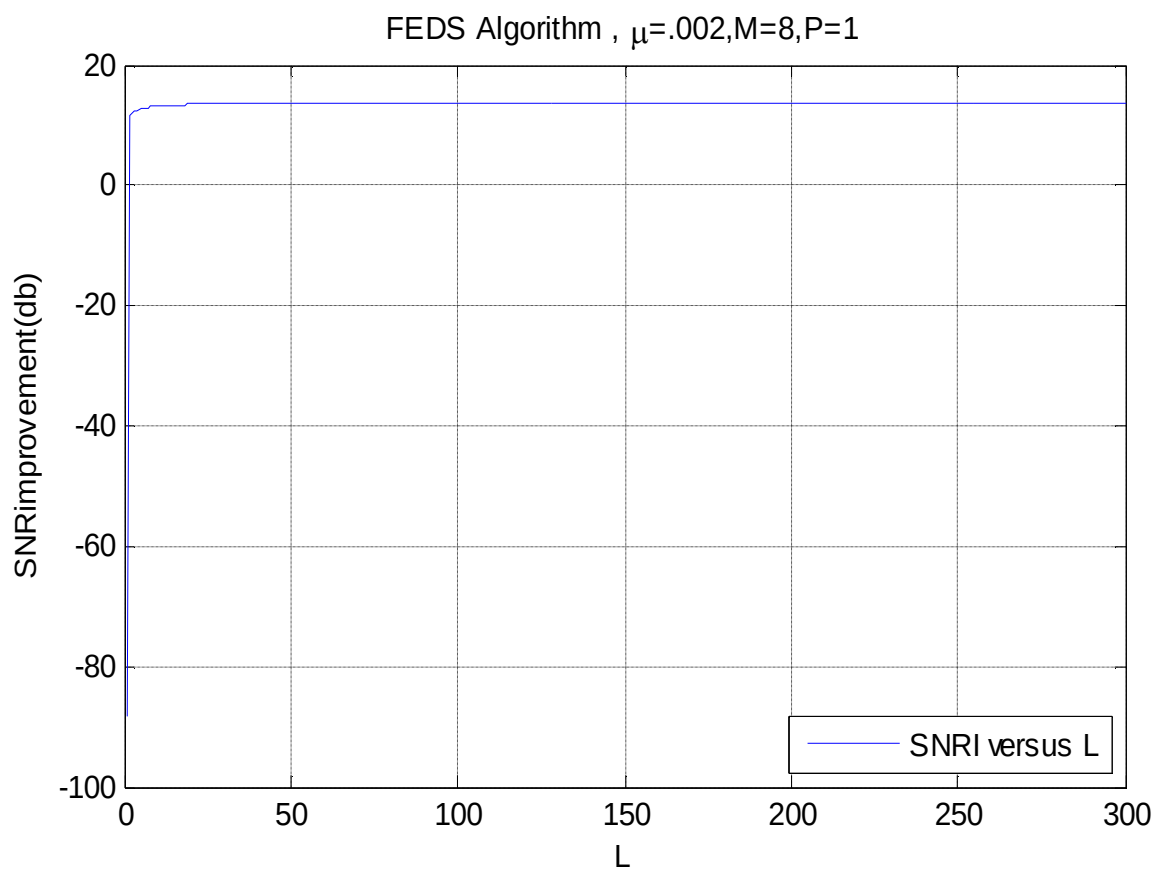

Fig. 18. SNRI versus L for FEDS algorithm.

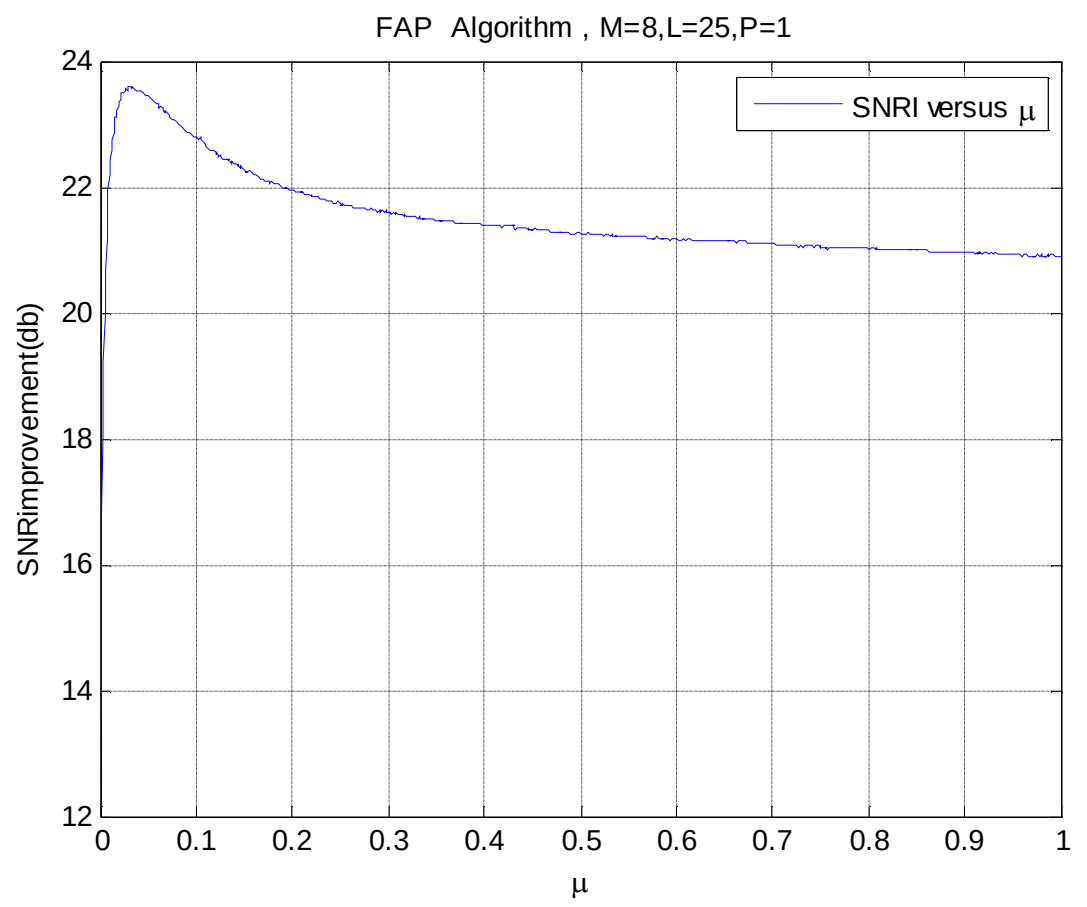

Fig. 21. SNRI versus μ for FAP algorithm.

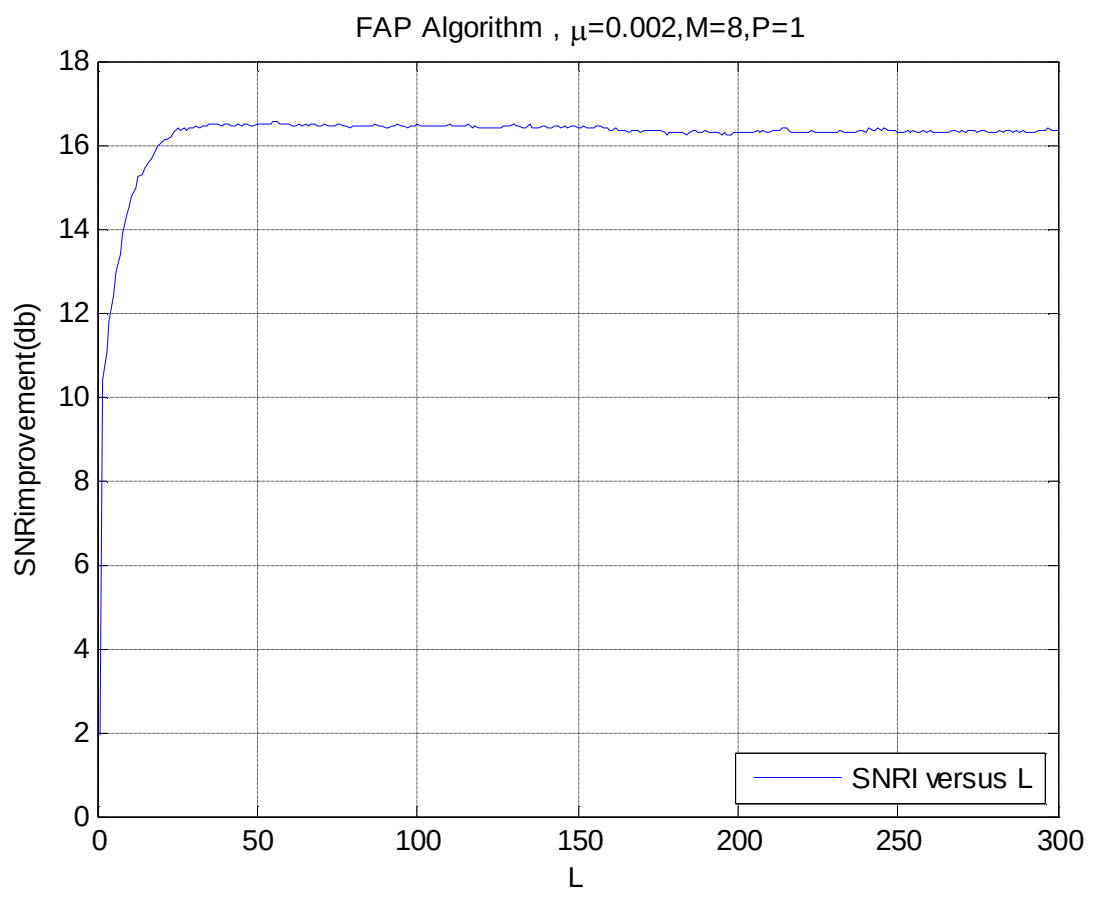

Fig. 19. SNRI versus L for FAP algorithm

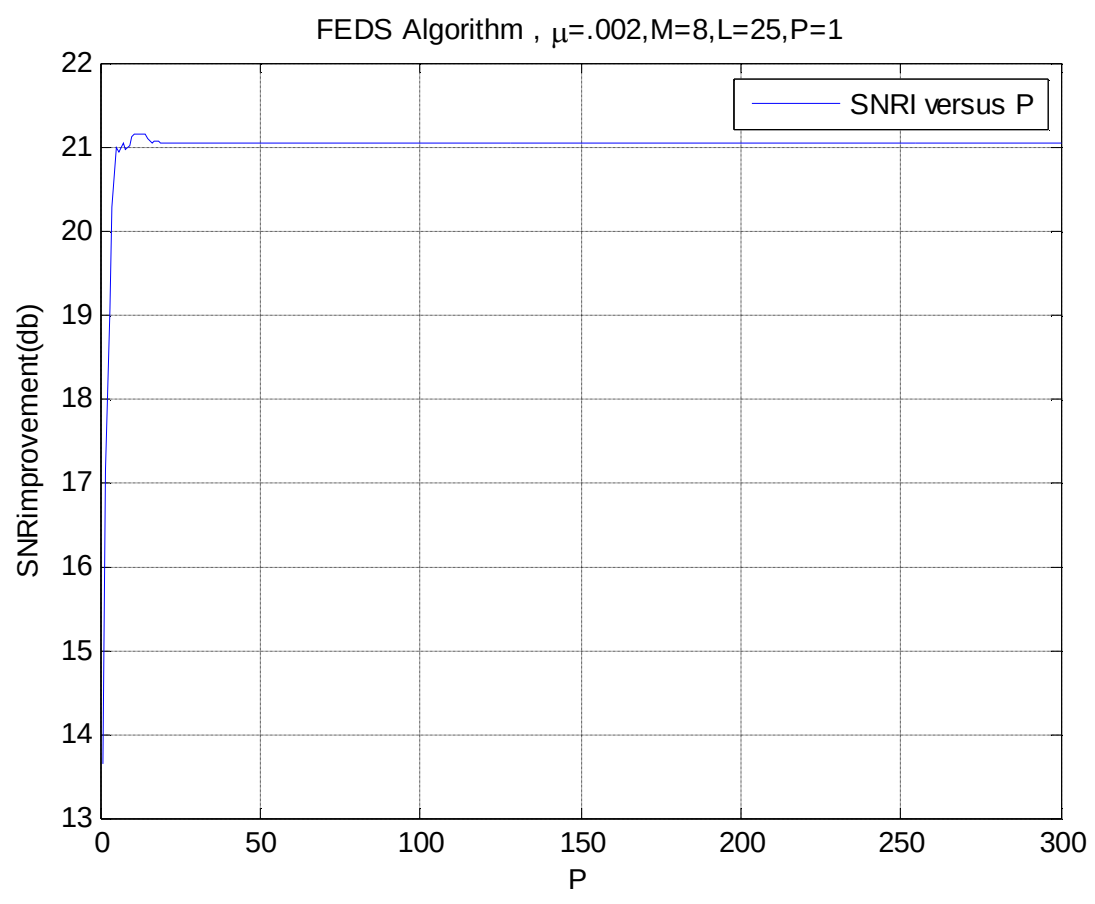

Fig.22. SNRI versus P for FEDS algorithm.

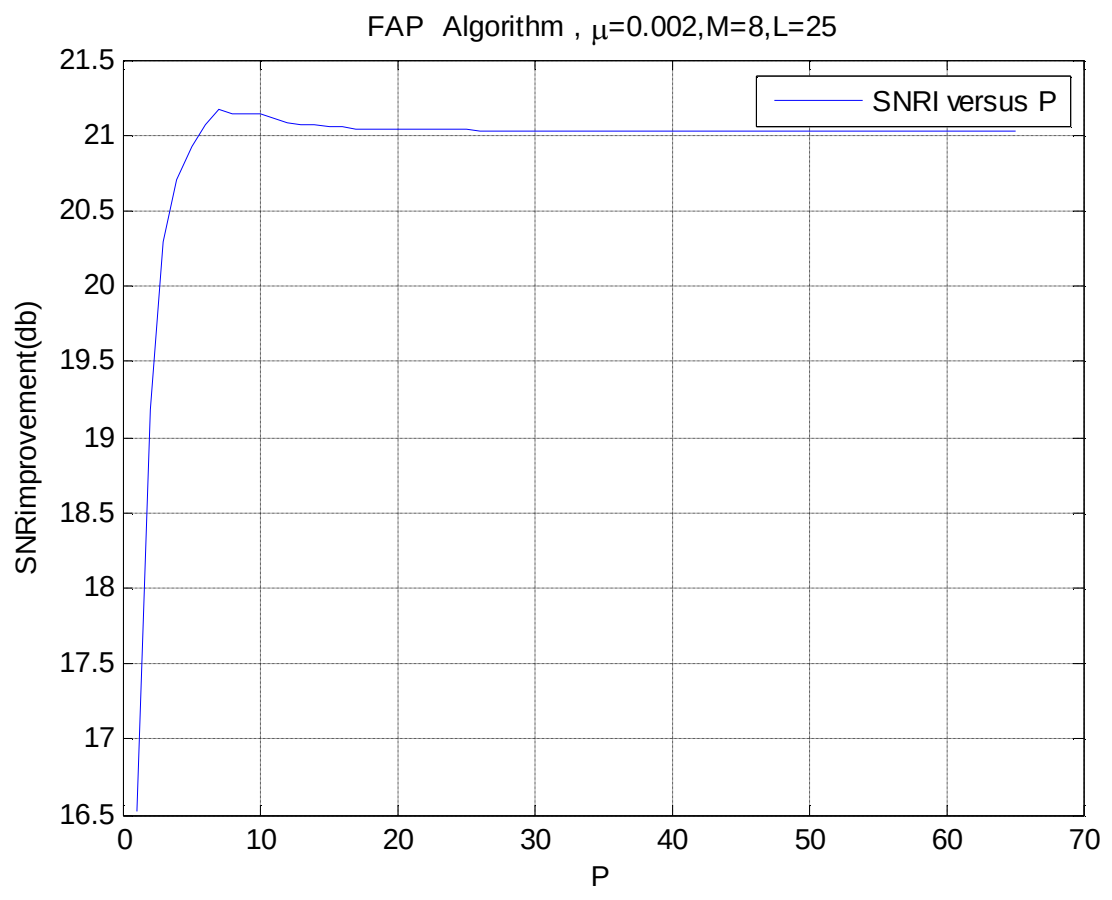

Fig.23. SNRI versus P for FAP algorithm.

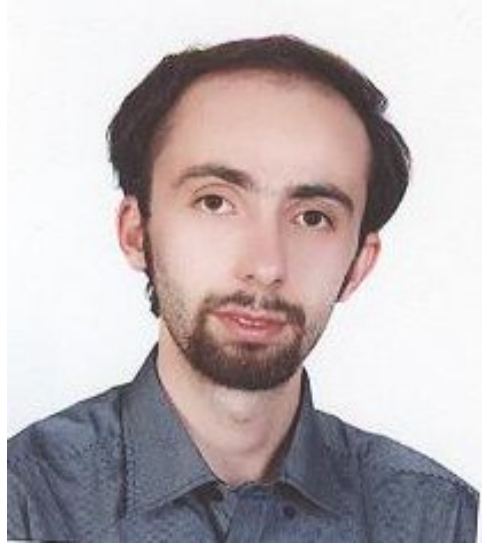

**Sayed.A. Hadei** was born in Ardebil, Iran in 1985. He has been working towards M.Sc degree in Electrical Engineering emphais on communication Systems from Tarbiat Modares University Tehran, Iran. His current research interest include digital filter theory, adaptive signal processing algorithms, bayesian signal processing, wireless communication, MIMO-OFDM systems and estimation theory.

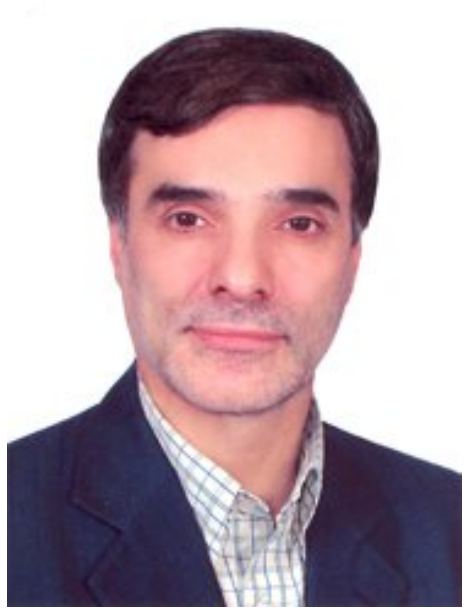

**Mojtaba Lotfizad** was born in Tehran, Iran, in 1955. He received the B.S. degree in electrical engineering from Amir Kabir University of Iran in 1980 and the M.S. and Ph.D.degrees from the University of Wales, UK, in 1985 and 1988, respectively. He joined the engineering faculty of Tarbiat Modarres University of Iran. He has also been a Consultant to several industrial and government organizations. His current research interests are signal processing, adaptive filtering, and speech processing and specialized processors.